\documentclass[12pt,a4paper]{article}
\usepackage[latin1] {inputenc}
\usepackage[T1]{fontenc}
\usepackage{amsmath}
\usepackage{amsfonts}
\usepackage[english]{babel}
\usepackage[dvips]{graphicx,color}
\begin{document}
\title{Spin in the Worldline Path Integral in $2+1$
Dimensions}
\author{C.~D.~Fosco$^{a}$\,,
 J.~S\'anchez-Guill\'en$^{b}$ and
  R.~A.~V\'azquez$^{b}$\\
  {\normalsize\it $^a$Centro At\'omico Bariloche and Instituto Balseiro}\\
  {\normalsize\it Comisi\'on Nacional de Energ\'{\i}a At\'omica}
  \\
  {\normalsize\it 8400 Bariloche, Argentina.}\\
  {\normalsize\it $^b$Departamento de F\'{\i}sica de  Part\'{\i}culas}\\
  {\normalsize\it Facultad de F\'{\i}sica and Instituto Galego de F\'{\i}sica 
  de Altas Energ\'{\i}as} \\
  {\normalsize\it Universidad de Santiago}\\
  {\normalsize\it E-15782 Santiago de Compostela, Spain.} }
\date{}
\maketitle
\begin{abstract}
We present a constructive derivation of a worldline path integral for the
effective action and the propagator of a Dirac field in $2+1$ dimensions,
in terms of spacetime and $SU(2)$ paths.  
After studying some general properties of this representation,
we show that the auxiliary gauge group variable can be integrated, 
deriving a worldline action depending only on $x(\tau)$, the spacetime
paths. 

\noindent We then show that the functional integral automatically
imposes the constraint $\dot{x}^2(\tau) = 1$, while there is a spin
action, which agrees with the one one should expect for a spin-$\frac{1}{2}$
field.
\end{abstract}
\section{Introduction}
The spacetime picture for (quantum) spin degrees of freedom and, 
in particular, of their classical limit, has been a subject of intense 
research, because of its many interesting subtleties, 
both from the physical and mathematical points of view.

Phase-space formulations for spin degrees of freedom which, when quantized,
lead to the proper dynamics of spinning particles, have been known since some
time ago~\cite{bala}. Those approaches introduce dynamical variables which are
group elements, a fact that naturally renders their path-integral formulation far
from trivial~\cite{orla} (although they avoid the use of Grassmannian
variables).
The starting point for the known constructions is usually algebraic
or geometric. Although, in the end, the usual propagator for a relativistic
spinning particle can be recovered, we believe that a more direct approach to
the problem is possible: to construct a quantum description for the spin degrees
of freedom starting from the knowledge of the propagator for a
Dirac field. As a concrete step in that direction, we shall do that here for the
case of a Dirac field in $2+1$ dimensions, in the worldline
approach~\cite{Feynman,Schwinger}, where quantum fluctuations are represented as
spacetime trajectories in an auxiliary, {\it proper time}, variable. After some
changes of variables within the worldline path integral, we shall obtain a
spacetime picture where the geometric interpretation is an emergent property,
rather than an \emph{ab initio} ansatz. 

We note that the handling of the spin degrees of freedom in this context,
is usually achieved by means of the introduction of Grassmann
variables~\cite{Bern,Strassler,Alexandrou}. Other methods, which provide a
physically more appealing picture of the spinning degrees of freedom have also
been considered, like the coherent-state path
integrals~\cite{Stone}, a historical review of which can be found
in~\cite{Frydryszak}. 

In this paper, we shall show that one can, indeed, construct a worldline
path integral at least in $2+1$ dimensions involving only $c$-number, commuting
variables. This preserves the intuitive appeal of the worldline
representation, and at the same time has the interesting practical
advantage that it is, in principle, much easier to deal with numerically
(i.e., on a lattice).  We shall derive that  worldline representation by
following a {\emph constructive\/} approach: as a starting point, we use an old
proposal, originally due to Migdal~\cite{Migdal} and which has been
extensively tested recently~\cite{us}. That first-order (phase space) path
integral is then transformed into an equivalent one in terms of spacetime
paths $x_\mu(\tau)$ and also $g(\tau)$ paths where \mbox{$g(\tau) \in
SU(2)$}, introduced by means of an alternative geometric
parametrization for the original integral over the canonical
momentum variable.  

This `intermediate' worldline representation is
studied, in a semiclassical expansion, to shed light on the dynamics of the
spin degrees of freedom, both for the free case and for a system in an
external gauge field background. We show that the classical equations of
motion can be written in terms of the intrinsic geometric properties of 
the spacetime path (arc length, curvature and torsion). Besides, by
calculating the leading contribution to the path integral in that
expansion, we show that the modulus of $\dot x(\tau)$ should be constant
and equal to $1$, a result which is later on shown to be exact at the
quantum level.

Finally, the transformation properties of the gauge group variables are
used in order to evaluate the gauge group path integral, what yields a
worldline path integral over spacetime paths only, weighed by an `effective
worldline action', which selects paths with $\dot x^2(\tau)= 1$ and
attaching to them a phase given by a spin action. 

The organization of this paper is as follows: in section~\ref{sec:method},
we first review the main properties of the first-order worldline construction.
Then, based on that construction,  we introduce the gauge group path
integral representation, studying the saddle point properties of the path
integral over the group variables.
In section~\ref{sec:consequences}, we use that representation to derive a few
results. In particular, we calculate the path integral over the group
variables to derive the effective worldline action, and afterwards we
evaluate the free propagator.

In Appendix A, we review an alternative representation for the integral
over the canonical momentum, which uses Grassmann variables to deal
with the matrix structure of the path ordered exponential. It provides an
equivalent way of treating the system that does not require the explicit
introduction of matrices.  We recall it here in order to compare it with 
the results obtained with the gauge group approach.

\section{The method}\label{sec:method}
In this section we shall introduce a novel representation for the effective
action and the propagator for a Dirac field in the presence of an external
gauge field. Both will involve a sum over two different kinds of paths:
$x_\mu(\tau)$, the spacetime paths, and $g(\tau)$, paths of an $SU(2)$ `gauge
group field', related to the evolution of `internal' degrees of freedom. 
Afterwards (see section~\ref{sec:consequences}), we will show that it is
possible to carry out the integration over the auxiliary gauge group
variables.

To begin with, we set up our conventions, and briefly review the main
features of the first-order formalism, which is used as starting point for
further developments. The action $S_f$ for a Dirac field in an Abelian
gauge field background $A_\mu$, in $2+1$ Euclidean dimensions has the form:
\begin{equation}\label{eq:defsf}
S_f({\bar\psi},\psi,A) \;=\; \int d^3x \, {\bar\psi} (\not\!\! D +
m) \psi \;,
\end{equation}
where
\begin{equation}\label{eq:dirconv}
\not\!\! D \equiv \gamma_\mu D_\mu ,
\; D_\mu = \partial_\mu + i e A_\mu \;,
\; \gamma_\mu^\dagger
= \gamma_\mu,\;\mu = 1,2,3 \;,
\end{equation}
and $e$ is the coupling constant. The Dirac matrices are chosen to be \mbox{$\gamma_\mu =
\sigma_\mu$}, \mbox{$\mu=1,\,2,\,3$}, where $\sigma_\mu$ denote the usual Pauli 
matrices.

Our starting point for the subsequent construction shall be a first-order
worldline path integral, whereby the one-loop effective action $\Gamma_f(A)$
(normalized to $\Gamma_f(0)= 0$):
\begin{eqnarray}
\Gamma_f(A) & \equiv\; & - \ln \Big[ \frac{\det \big(\not \!\! D + m
  \big)}{\det 
  \big(\not \! \partial + m \big)} \Big] \nonumber  \\ 
 & = &  - {\rm Tr} \ln (\not \!\! D + m \big) \;+\;
{\rm Tr} \ln  (\not \! \partial + m \big),
\label{eq:defga}
\end{eqnarray}
is represented as~\cite{Migdal}:
\begin{eqnarray}\label{eq:gppa}
\Gamma_f (A) & = &  \int_{0}^\infty \frac{dT}{T} \, e^{- m T} \,
\int_{x(0)=x(T)} {\mathcal D}x \,  {\mathcal D}p \, e^{ i \int_0^T d\tau
 p_\mu (\tau){\dot x}_\mu (\tau) } \nonumber \\
& \times &   {\rm tr}\big[ {\mathcal P} e^{-i \int_0^T d\tau \not
    p(\tau)} \big] 
\; \Big[ e^{-i e \int_0^T d\tau {\dot x}_\mu(\tau) A_\mu[x(\tau)] }- 1
\big] \;.
\end{eqnarray}
The fermion propagator, $G_f(x,y)$, is given by a similar integral, albeit
without the Dirac trace and with different boundary conditions: $x(T)=x$,
$x(0)=y$: 
\begin{eqnarray}
G_f(x,y) & = &  \int_{0}^\infty dT \, e^{- m T} \,
\int_{x(0)=y}^{x(T)=x} {\mathcal D}x \,  {\mathcal D}p \, e^{ i \int_0^T d\tau
 p_\mu {\dot x}_\mu } \nonumber \\
& \times &  \Big[ {\mathcal P} e^{-i \int_0^T d\tau \not p(\tau)} \big]
\; e^{-i e \int_0^T d\tau {\dot x}_\mu(\tau) A_\mu[x(\tau)] }\;.
\label{eq:gfppa}
\end{eqnarray}
A path integral measure for $\Gamma_f(A)$ that takes
into account the boundary conditions, can be formally written as: 
\begin{equation}
\Big[ {\mathcal D}x \, {\mathcal D}p \Big]_\Gamma
\;\equiv\;
d^3x(0) \, 
\Big[ \prod_{0 < \tau \leq T} \, \frac{ d^3x(\tau) \, d^3p(\tau)}{(2\pi)^3} 
\Big] \; \delta[x(T) - x(0)] \;,   
\end{equation}
where the $\delta$ function, of course, imposes the periodicity condition.
Note that the measure is dimensionless (in $\hbar=1$ units), as it should
be.
A more symmetric expression may also be written for this measure,
\begin{equation}
\Big[ {\mathcal D}x \, {\mathcal D}p \Big]_\Gamma
\;\equiv\;
\Big[\prod_{0 \leq \tau \leq T} \, \frac{ d^3x(\tau) \, d^3p(\tau)}{(2\pi)^3} 
\Big] \, \delta[x(T) - x(0)] \, \delta[p(T) - p(0)] \;,   
\end{equation}
where the extra integration over $p(0)$ is harmless, since it yields $1$
because of the new delta function. Note that this means that, if the
integration ranges for both phase space variables are equal, then there is
periodicity in both of them.

For the propagator, we have instead:
\begin{equation}
\Big[{\mathcal D}x \,  {\mathcal D}p \Big]_G \;\equiv\;
\frac{d^3p(0)}{(2\pi)^3} 
\prod_{0 < \tau < T} \, \frac{ d^3x(\tau) \, d^3p(\tau)}{(2\pi)^3} \;,   
\end{equation}
which has the dimensions of a $({\rm mass})^3$, which combined with the
$({\rm mass})^{-1}$ of the $dT$ factor, yields the proper mass dimensions
to the fermion propagator in coordinate space. In what follows, we shall
omit writing explicitly the suffix ($\Gamma$ or $G$) to  identify the
measure, since that shall be clear from  the context. 

Since the following manipulations will only involve the $p_\mu(\tau)$
integrals, we formally disentangle them from the $x_\mu(\tau)$ integrals by
introducing the definition:
\begin{eqnarray}
\Gamma_f (A) & = &  \int_{0}^\infty \frac{dT}{T} \, e^{- m T} \,
\int_{x(0)=x(T)} {\mathcal D}x \nonumber \\ & \times & 
e^{- S[x(\tau)]} \; \Big\{ e^{- i e \int_0^T d\tau
{\dot x}_\mu(\tau) A_\mu[x(\tau)]} \,-\,1 \Big\}\; ,
\label{eq:gppa1}
\end{eqnarray}
and similarly for the propagator:
\begin{eqnarray}
G_f(x,y) & = &  \int_{0}^\infty dT \, e^{- m T} \,
\int_{x(0)=y}^{x(T)=x} {\mathcal D}x \nonumber \\ & \times & 
D[x(\tau)] \; e^ {-i e \int_0^T d\tau {\dot x}_\mu(\tau) A_\mu[x(\tau)] },
\label{eq:gfppa1}
\end{eqnarray}
where $S[x(\tau)]$ and $D[x(\tau)]$ are defined by the integrals:
\begin{equation}\label{eq:defsefx}
 e^{- S[x(\tau)]} = \int {\mathcal D}p \,
{\rm tr}\big[ {\mathcal P} e^{-i \int_0^T d\tau \not p(\tau)} \big] 
e^{ i \int_0^T d\tau p_\mu(\tau) {\dot x}_\mu(\tau) } \;,
\end{equation}
and
\begin{equation}\label{eq:defdefx}
 D[x(\tau)] \;=\; \int {\mathcal D}p \,
\Big[ {\mathcal P} e^{-i \int_0^T d\tau \not p(\tau)} \Big] \; 
e^{ i \int_0^T d\tau p_\mu(\tau) {\dot x}_\mu(\tau) } \;.
\end{equation}
Of course, the two previously defined objects are related by:
\begin{equation}
 e^{- S[x(\tau)]} = {\rm tr}\Big\{ D[x(\tau)]\Big\} \;.
\end{equation}

From their definitions, these scalar and matrix functionals 
(resp. (\ref{eq:defsefx}) and (\ref{eq:defdefx}))
can be regarded as {\em functional\/} Fourier transformations of two particular
functionals of $p(\tau)$. Since these functionals are neither Gaussian, nor
do they have a simple structure, it is not evident at all how to evaluate
their Fourier transformations. In order to tackle that problem, we shall
introduce a change of variables in the $p(\tau)$ integral which shall
render their integration easier. At the same time, some features of the
spin dynamics will become more transparent.

To prepare the ground for that change of variables, we find it convenient
to introduce first some notation and conventions. First, we define the
(anti-Hermitian) matrices: $\lambda_\mu \equiv \frac{\sigma_\mu}{2 i}$,
which can be regarded as a basis for $su(2)$, the Lie algebra of $SU(2)$~\cite{rubakov}. 

Using that basis, for each value of $\tau$ we associate to $x_\mu(\tau)$ and $p_\mu(\tau)$ two
(anti-Hermitian) elements in $su(2)$: $\chi(\tau)$ and $\pi(\tau)$,
respectively, according to the rules: 
\begin{eqnarray}\label{eq:defalg}
\chi(\tau) &=& \lambda_\mu \, \chi_\mu(\tau) \;,\;\;\; \chi_\mu(\tau)
\;=\;x_\mu(\tau) \nonumber\\  
\pi(\tau) &=& \lambda_\mu \, \pi_\mu(\tau)\;,\;\;\; \pi_\mu(\tau)\;=\;
2 \, p_\mu(\tau) \;.
\end{eqnarray}

With this notation, $S[x(\tau)]$ adopts a slightly 
more compact form:
\begin{eqnarray}
 e^{- S[x(\tau)]} &=&  \det(\frac{\delta_{\mu\nu}}{2} I ) \, \int\,
 {\mathcal D}\pi \; 
{\rm tr}\big[ {\mathcal P} e^{ \int_0^T d\tau \, \pi(\tau)} \big] 
\nonumber \\
& \times & e^{- i \int_0^T d\tau {\rm tr}[ \pi(\tau) {\dot \chi}
(\tau)] }\;,
\label{eq:sef1}
\end{eqnarray}
where $I$ is the identity operator acting on periodic functions from
$[0,T]$. $\det(\frac{\delta_{\mu\nu}}{2} I)$ accounts for the Jacobian
under the change $p_\mu \to \pi_\mu$ which is an infinite constant. Of
course, the corresponding expression can be constructed for $D[x(\tau)]$,
with the only change of omitting the trace.
\subsection{Gauge group variables}
To proceed, we introduce a parametrization for the integral over
$\pi(\tau)$, in terms of gauge group variables. The crucial point here is
to note that $\pi(\tau)$ may in fact be regarded as an $SU(2)$ gauge field
defined on a $0+1$-dimensional `spacetime' (with $\tau$ as the time).  We
next introduce gauge transformations for this gauge field, which adopt the form:
\begin{equation}
\pi(\tau) \;\to\; \pi^g(\tau) \,\equiv\, 
g(\tau) \pi(\tau) g^{-1}(\tau) \,+\,  g(\tau) \partial_\tau g^{-1}(\tau),
\end{equation}
where $g(\tau) \in SU(2)$. It should be clear that neither (\ref{eq:sef1})
nor the corresponding expression  for the propagator are invariant under
these transformations, and that they bear no relationship with the
gauge symmetry due to coupling to an external gauge field. 

Nevertheless, we can still use the Faddeev-Popov (FP) trick of inserting a
$1$ inside the path integral, corresponding to a gauge fixing for those
transformations.  Of course, since this is not a gauge-invariant theory,
the gauge group integration shall not factorize out of the integral, and
the group variable will become dynamical. 

The usual expression for the FP `$1$' becomes, in this case: 
\begin{equation}\label{eq:fp1}
1 \;=\; \int {\mathcal D}g \, \delta[{\mathcal F}(\pi^g)] \;
\Delta_{\mathcal F}[\pi], 
\end{equation}
where ${\mathcal F}(\pi)= ({\mathcal F}_\mu(\pi))_{\mu=1}^3$ is the
gauge-fixing functional (with three components, its coordinates in the
Lie algebra) and $\Delta_{\mathcal F}[\pi]$ its corresponding FP
(gauge-invariant) determinant.

Including this $1$ into the expression leading to $S[x(\tau)]$, we see
that:
\begin{eqnarray}
 e^{- S[x(\tau)]} & = & \det(\frac{\delta_{\mu\nu}}{2} I) \, \int\, 
{\mathcal D}g \, {\mathcal D}\pi \;
 \delta[{\mathcal F}(\pi^g)] \nonumber \\  
&\times &
\Delta_{\mathcal F}[\pi] \; 
{\rm tr}\big[ {\mathcal P} e^{ \int_0^T d\tau \, \pi(\tau)} \big] 
\nonumber \\
& \times & e^{- i \int_0^T d\tau {\rm tr}[\pi(\tau) {\dot \chi}(\tau)] }\;.
\label{eq:sef2}
\end{eqnarray}

Performing now the change of variables: $\pi \to \pi^{g^{-1}}$ in
(\ref{eq:sef2}) leads to:
\begin{eqnarray}
 e^{- S[x(\tau)]} & = &  \det(\frac{\delta_{\mu\nu}}{2} I) \, \int\, 
{\mathcal D}g \, {\mathcal D}\pi \;
 \delta[{\mathcal F}(\pi)] \; \Delta_{\mathcal F}[\pi] \nonumber \\
& \times & {\rm tr}\big[g^{-1}(T) {\mathcal P} e^{ \int_0^T d\tau \, 
\pi(\tau)} g(0)\big] \nonumber \\
& \times & e^{- i \int_0^T d\tau {\rm tr}[\pi^{g^{-1}}(\tau) {\dot
      \chi}(\tau)]}\;.
\label{eq:sef3}
\end{eqnarray}
We have taken into account the properties of invariance under gauge
transformations both for the $\pi(\tau)$ measure and the FP determinant,
and also the fact that, under the transformation \mbox{$\pi(\tau) \to
\pi^{g^{-1}}(\tau)$}, the path-ordered factor behaves as a Wilson line:
\begin{equation}
{\mathcal P} e^{ \int_0^T d\tau \, \pi(\tau)}\;\to\; 
g^{-1}(T) {\mathcal P} e^{ \int_0^T d\tau \, \pi(\tau)} g(0) \;.
\end{equation}
To proceed, we must decide which particular gauge fixing ${\mathcal F}$ to
use.  Since we only have one spacetime component in $\pi(\tau)$ (note that
the $\pi_\mu(\tau)$ are `internal space' components from the point of view
of the $\tau$-spacetime), the simplest possible choice is of course:
\begin{equation}\label{eq:gf}
{\mathcal F}_\mu \;=\; \pi_\mu \;,
\end{equation}
a `temporal gauge', which in this case will erase the gauge field
completely, at the expense of introducing $g(\tau)$ as a dynamical
variable, since the integrand is not gauge-invariant. Indeed, using
(\ref{eq:gf}) into (\ref{eq:sef3}), we see that:
\begin{eqnarray}
 e^{- S[x(\tau)]} & = & \det(\frac{\delta_{\mu\nu}}{4 \pi}I)
 \,\det(\partial_\tau) \; \int\,  
{\mathcal D}g \, {\rm tr}\big[g^{-1}(T) \, g(0)\big] \nonumber \\
& \times &  e^{- i \int_0^T d\tau {\rm tr}[g^{-1}(\tau) \partial_\tau
 g(\tau) {\dot \chi}(\tau)]}\;, 
\label{eq:sef4}
\end{eqnarray}
where $\det(\partial_\tau)$ is what remains of the FP determinant when it
is evaluated on the gauge-fixed slice $\pi_\mu(\tau)=0$, and we included
the $(2\pi)^{-1}$ factors that came with the definition of the
integration measure.

Thus we have obtained the `intermediate' representation (\ref{eq:sef4}) for
$S$, where no path-ordering appears, and the integration variables are
$x_\mu(\tau)$ and $g(\tau)$. The two $x_\mu(\tau)$-independent factors may
be included into a constant ${\mathcal N}$ so that we shall write:
\begin{eqnarray}\label{eq:sef5}
 e^{- S[x(\tau)]} &=& 
\frac{1}{\mathcal N} \; \int\,{\mathcal D}g \; 
{\rm tr}\big[g^{-1}(T) \, g(0)\big] \nonumber \\
& \times &  e^{- i \int_0^T d\tau {\rm tr}[g^{-1}(\tau)\partial_\tau 
g(\tau) {\dot \chi}(\tau)]} \;.
\end{eqnarray}

\subsection{Classical limit}\label{ssec:semiclassical}
In order to interpret the path integral over $g(\tau)$, leading to the
action $S[x(\tau)]$, we shall first consider its saddle point equation, as
derived from the leading term in the corresponding semiclassical
expansion. Even at this stage, a non trivial condition on the paths shall
arise, namely, that the modulus of $\dot x(\tau)$ has to be constant at
the saddle point. Remarkably, the same condition will emerge at the
quantum level, as an exact constraint.

Starting from the defining expression for $S$:
\begin{eqnarray}
 e^{- S[x(\tau)]} & = & \frac{1}{\mathcal N} \;
\int\,  
{\mathcal D}g \, {\rm tr}\big[g^{-1}(T) \, g(0)\big] \nonumber \\
& \times &  \exp\big\{- \frac{1}{\hbar} \, I[l,\chi] \big\}\;, 
\end{eqnarray}
where we have reinserted an $\hbar$ factor, and we defined the
functional:
\begin{equation}
I[l,\chi] \equiv i \int_0^T d\tau {\rm tr}[l(\tau) {\dot \chi}(\tau)]
\end{equation}
with
\begin{equation}
l(\tau) = \lambda_\mu l_\mu(\tau)= g^{-1}(\tau) \partial_\tau g(\tau) \;.
\end{equation}
To find the saddle points of the above expression with respect to $g$
($\chi$ is here to be regarded as `external' in the $g$ integral), we
consider variations with
respect to $g$. Since such variations should still leave the field in
$SU(2)$, we may parametrize them as follows:
\begin{equation}
g(\tau) \;\equiv\; {\tilde g}(\tau) h(\tau),
\end{equation}
where both ${\tilde g}(\tau)$ and $h(\tau)$ are in $SU(2)$. For
${\tilde g}(\tau)$ we assume it to verify the boundary conditions:
\mbox{${\tilde g}(0) = g(0)$} and \mbox{${\tilde g}(T) = g(T)$}, while
\mbox{$h(0) = h(T) = 1$}.

To find the extrema of the action, it is sufficient to consider
an $h(\tau)$ infinitesimally close to the identity:
\begin{equation}
h(\tau) \,\simeq \, 1 + \alpha(\tau), 
\end{equation}
where $\alpha(\tau)$ is an infinitesimal $su(2)$ matrix verifying 
\begin{equation}
\alpha(\tau)^{\dagger}= -\alpha(\tau),  \; 
\;\; {\rm tr} \, \alpha(\tau)\, =\, 0\; .
\end{equation}
Under this transformation $l$ behaves as follows:
\begin{equation}
l(\tau) \;\to\;  l(\tau) + \delta_\alpha l(\tau),
\end{equation}
where 
\begin{equation}
\delta_\alpha l(\tau) \,=\, D_\tau \alpha(\tau) \equiv 
\partial_\tau \alpha(\tau) + [l(\tau),\alpha(\tau)] \, .
\end{equation}
Then the saddle point equation follows from the requirement that 
\begin{equation}
\delta_\alpha \int_0^T d\tau \, {\rm tr} \big[ l(\tau) \dot \chi(\tau)
\big] = 0,
\end{equation}
which is equivalent, after an integration by parts, to: 
\begin{equation}
0 = - \int_0^T d\tau \, {\rm tr} \big[\alpha(\tau) D_\tau \dot
\chi(\tau)\big]\;.
\end{equation}
Then the differential equation for the extrema is:
\begin{equation}
D_\tau \dot \chi(\tau) = 0\;,
\end{equation}
a sort of constant covariance condition which, in components, reads as follows:
\begin{equation}
\ddot x_\mu(\tau) \,+\, \epsilon_{\mu \nu \rho} \, l_\nu(\tau)
\, \dot x_\rho(\tau) \;=\; 0 \;.
\end{equation}
We insist that the above equation must be thought of as an equation for
${\tilde g}$ (via ${\tilde l}$), and not for ${\dot x}_\mu$, which does
indeed appear in the equation, but  as an {\em external\/} function. 
However, we do get from the above equation a consistency condition
(required in order to have a saddle point). Indeed, contracting with 
${\dot x}_\mu$, we get:
\begin{equation}
\dot x \cdot \ddot x = 0 \Rightarrow \frac{d \dot x^2}{d\tau} = 0 \;,
\end{equation}
{\it i.e. } $\dot x^2 = v^2 = $ constant. This means that, {\em in
order for a semiclassical limit to exist}, the modulus of ${\dot
x}_\mu(\tau)$ has to be a constant, a condition that we shall assume is,
indeed, fulfilled.

In the geometric analysis that now follows, we find it convenient
to introduce the unit tangent vector $t_\mu \equiv \dot x_\mu/v$,
($t^2 = 1$), and to parametrize everything in terms of the arc length 
$s$, which is in fact proportional to $\tau$:
\begin{equation}
ds \equiv \sqrt{\dot x^2} d\tau = v d\tau \;\Rightarrow \; s \,=\, 
v \tau \,.
\end{equation}
Introducing the notation: $f'(s) \equiv \frac{d f}{ds}$, we then have:
\begin{equation}\label{eq:tp}
t'_\mu (s) + \epsilon_{\mu \nu \rho} {\tilde l}_\nu(s) t_\rho(s)
\,=\, 0 \;,
\end{equation}
where ${\tilde l}(s) \equiv {\tilde g}^{-1}(s) \partial_s  {\tilde
g}(s)$.
Before proceeding, we note that (\ref{eq:tp}) is exactly like the equation
for a `spin' variable~\cite{zinn}, $t_\mu$, in the presence of a
time-dependent `magnetic field' background ${\tilde l}_\mu$.  Note,
however, that it has to be regarded as an equation determining 
${\tilde l}_\mu(s)$.  We immediately see that that equation can only fix 
${\tilde l}_\mu(s)$ modulo a term proportional to $t_\mu (s)$. 

\noindent We shall then, for the sake of simplicity, assume that 
${\tilde l}_\mu (s)$ is orthogonal to $t_\mu$ in
(\ref{eq:tp}), keeping in mind that we can afterwards add to the solution
an arbitrary term proportional to the tangent vector.

By contracting (\ref{eq:tp}) with ${\tilde l}_\mu$, we also see that it is
also orthogonal to $t'_\mu$. In other words, ${\tilde l}$ is orthogonal to
$n$, 
where $n_\mu(s)$ is the principal normal to the curve
$x_\mu(s)$ at the point $s$. 

Since (by assumption) ${\tilde l}_\mu$ is orthogonal to $t_\mu$, we see that 
${\tilde l}_\mu$ must, in fact, have the direction of the {\em binormal\/} $b_\mu$, 
which is given by:
\begin{equation}\label{eq:defbinormal}
b_\mu (s) \;=\; \epsilon_{\mu\nu\rho} t_\nu(s) \, n_\rho(s) \;,
\end{equation} 
Thus,
\begin{equation}
{\tilde l}_\mu(s) = \eta (s) b_\mu (s)
\end{equation}
and the function $\eta (s)$ can be determined from the fact that
one of Frenet's equations is just the assertion that $t'(s)$ is equal
to the curvature $\kappa (s)$ times $n_\mu (s)$. Since, on the other
hand, $\epsilon_{\mu\nu\rho} b_\nu  t_\rho = n_\mu$, we see that the
equation of motion implies:
\begin{equation}
{\tilde l}_\mu (s) = - \kappa (s) b_\mu (s) \;, 
\end{equation}
and this determines completely the component of ${\tilde l}_\mu$ that is
orthogonal to $t_\mu(s)$. 
This may be put more explicitly in terms of the curve's equation, as
follows:
\begin{equation}
{\tilde l}_\mu (s) = - \epsilon_{\mu\nu\rho} x'_\nu(s) \, x''_\rho(s) \;, 
\end{equation}
or, in terms of $\tau$,
\begin{equation}
l_\mu (\tau) \,=\, - v^{-2} \, \epsilon_{\mu\nu\rho} \dot
x_\nu(\tau) \, \ddot x_\rho(\tau) \;.
\end{equation}

Of course, since it is orthogonal to $\dot x_\mu$ it yields no contribution
to the value of the functional $I$ at the saddle point. In other words, the
saddle-point action is independent the component of ${\tilde l}_\mu$ which
is in the direction of the binormal, although that component undergoes a
non trivial evolution as a result of that equation.

Then the only contribution comes from the `longitudinal' part,
${\tilde l}^\parallel$ proportional to $t_\mu$:
\begin{equation}
{\tilde l}_\mu^\parallel (\tau) \,=\, \xi (\tau) \,\dot x_\mu (\tau)\;.
\end{equation}
where $\xi$ parametrizes the arbitrariness related to the fact
that a term proportional to the tangent vector may be added to the solution
${\tilde l}$.

Then the saddle point contribution to the action $S[x(\tau)]$ is:
\begin{equation}
 e^{- S[x(\tau)]}| \; \sim \; \frac{1}{\mathcal N} \; \int\,  
{\mathcal D}\xi \, {\rm tr}\big[{\tilde g}^{-1}(T) \, {\tilde g}(0)\big]
\, \times \, e^{\frac{i}{2}  \int_0^T d\tau \xi (\tau) v^2  }\;. 
\end{equation}
Besides, ${\tilde g}$ can be of course determined from the
knowledge of ${\tilde l}$:
\begin{equation}
{\tilde g}^{-1}(L) \;=\; {\mathcal P}e^{ - \int_0^\tau d{\tilde\tau} l({\tilde\tau})} {\tilde g}^{-1}(0)
\end{equation}
so that:
\begin{equation}
{\rm tr}\big[{\tilde g}^{-1}(T) \, {\tilde g}(0)\big]
\;=\;
{\rm tr}\big[{\mathcal P} e^{ \int_0^T d\tau [ v^{-2} \epsilon_{\mu\nu\rho} 
\dot x_\nu \ddot x_\rho - \xi (\tau) \dot x_\mu(\tau) ] \lambda_\mu}\big] \;.
\end{equation}
Thus, the saddle point contribution involves an integration over $\xi$ (it
is not determined by the equations) and it adopts the form:
\begin{equation}
e^{- S[x(\tau)]} \;\propto \; \int {\mathcal D}\xi \, 
 {\rm tr}\big[{\mathcal P} e^{ \int_0^T d\tau [ v^{-2} \epsilon_{\mu\nu\rho} 
\dot x_\nu \ddot x_\rho - \xi (\tau) \dot x_\mu(\tau) ] \lambda_\mu}\big]
 e^{\frac{i}{2}  \int_0^T d\tau \xi (\tau) v^2  }\;.
\end{equation}

Let us now perform the exact integration over the `moduli' $\xi$; we shall
see that this yields a constraint on the constant $v$, the modulus of $\dot
x(\tau)$. 
The functional integral over $\xi$ can be discretized into $N$ ($N \to
\infty$) normal integrals, and at each discrete time $\tau_i$ we have to
evaluate the (normal) integral:
\begin{equation}
{\mathcal I}(\tau_i)\,\equiv\, \int_{-\infty}^{+\infty} \frac{d\xi(\tau_i)}{2\pi} e^{ \frac{i}{2} \delta \tau 
\xi (\tau_i) \big( \dot x_\mu(\tau_i) \sigma_\mu + v^2\big) }\;,  
\end{equation}
where $\delta \tau = T/N$. Taking into account the fact that:
\begin{eqnarray}
e^{ \frac{i}{2} \delta \tau \xi (\tau_i) \dot x_\mu(\tau_i) \sigma_\mu}
&=& \frac{1 + \sigma_\mu t_\mu (\tau_i)}{2} \, 
e^{ \frac{i}{2} \delta \tau \xi (\tau_i) }  \nonumber\\
&+& \frac{1 - \sigma_\mu t_\mu (\tau_i)}{2} \, 
e^{-\frac{i}{2} \delta \tau \xi (\tau_i) } \;,
\end{eqnarray}
we see that the integral over $\xi(\tau_i)$ yields $\delta$ functions:
\begin{equation}
{\mathcal I}(\tau_i)\,=\,\frac{2}{\delta \tau}\,\delta(v^2 -1) \,\frac{1 +
\sigma_\mu t_\mu (\tau_i)}{2} + \frac{2}{\delta\tau}\,\delta(v^2 + 1)
\,\frac{1 - \sigma_\mu t_\mu(\tau_i)}{2} \;.
\end{equation}
Since $v > 0$, only the first term survives and coming back to the
original expression for the leading semiclassical contribution, we see
that:
\begin{equation}
e^{- S[x(\tau)]} \propto \delta (v - 1)
{\rm tr}\Big\{{\mathcal P} \big[ e^{ \int_0^T d\tau 
\epsilon_{\mu\nu\rho} \dot x_\nu \ddot x_\rho \lambda_\mu}
\displaystyle{(\frac{1 + \sigma_\mu t_\mu (\tau)}{2})}
\big]
\Big\},
\end{equation}
or
\begin{equation}
e^{- S[x(\tau)]} \propto \delta (v - 1)\,
e^{-\frac{i}{2}\int_0^T d\tau \kappa(\tau)} 
{\rm tr} \Big\{ {\mathcal P} \big[(\frac{1 + \sigma_\mu b_\mu
(\tau)}{2}) (\frac{1 + \sigma_\mu t_\mu (\tau)}{2}) \big] \Big\},
\end{equation}
and
\begin{equation}
e^{- S[x(\tau)]} \propto \delta (v - 1)\,
e^{-\frac{i}{2}\int_0^T d\tau \kappa(\tau)} 
{\rm tr} 
\Big\{ {\mathcal P} \frac{1 + i \sigma_\mu n_\mu(\tau)}{2} \Big\}. 
\end{equation}

Let us finally consider what happens when the external gauge field
is included, so that we also have the equations of motion for $x$, as 
determined from its electromagnetic coupling:

\begin{equation}
\frac{\delta}{\delta x_\mu(\tau)} \int_0^T d\tau {\rm tr}[l(\tau) \dot
  \chi(\tau)] + e \, \dot x_\mu(\tau) A_\mu(x(\tau)) = 0,
\end{equation}
which implies 
\begin{equation}
\frac{1}{2} \dot l_\mu(\tau) = - \epsilon_{\mu\nu\rho} \, e \tilde F_\nu 
\, \dot x_\rho, 
\label{ldot}
\end{equation}
where we introduced the dual of $F_{\mu\nu}$, $\tilde F_\mu =
\epsilon_{\mu\nu\rho} F_{\nu\rho}$. 
We also see that 
\begin{equation}
\dot t \cdot l = 0, \;\;\;\; \dot l \cdot \tilde F =0, \;\;\;\;\; \dot l
\cdot t = 0.
\end{equation}
which implies that $t\cdot l$ = constant. 

For simplicity let us consider the free case, $e=0$,  Eq. (\ref{ldot}) 
implies $l_\mu = $ constant, and then Eq. (\ref{eq:tp}) gives
\begin{equation}
t_\mu(\tau) = (e^{-\tau L})_{\mu\nu} \; t_\nu(0),
\end{equation}
where 
\begin{equation}
L_{\mu\nu}(\tau) \equiv \epsilon_{\mu \nu\rho} \; l_\rho(\tau).
\end{equation}
by choosing $l$ to point in the 3 direction $l_\mu(0) = l \delta_{\mu
  3}$ we arrive at
\begin{equation}
\dot t_{3} = 0, 
\end{equation}
and 
\begin{equation}
\dot t_i = l \; \epsilon_{ij} \; t_j, \;\;\;\; i,j=1,2, 
\end{equation}
which gives as a solution
\begin{equation}
t_3 = a = {\rm const.}, \;\;\;\; t_1 = b \cos(|l|\tau + c) \;\;\;\; t_2 =
- b \sin(|l|\tau + c), 
\end{equation}
where $a$ and $b$ are constants, $a^2+b^2=1$, and $c$ is an arbitrary phase.
{\it i.e.} the tangent vector precesses even in the free (no field)
case. In coordinate space, this means that the particle describes an
helix. 

In the general case we have the coupled system of equations
(\ref{eq:tp}) and (\ref{ldot}) and the additional conditions 
\begin{equation}
\dot x^2 = {\rm constant}, \;\;\;\; t \cdot l = {\rm constant}. 
\end{equation}
\section{Consequences}\label{sec:consequences}
\subsection{Calculation of $S[x(\tau)]$}
To find $S[x(\tau)]$ is tantamount to obtaining a representation where only
$x_\mu(\tau)$ paths remain (with $S[x(\tau)]$ plus the Maxwell interaction 
term with $A$ as the weight).

We first note that there is an important Ward identity that follows by
performing an {\em infinitesimal} change of variables in the original
expression for $S$. Performing an infinitesimal change of variables 
$g(\tau) \, \to \, g(\tau) \,h(\tau)$ where $h(\tau)\simeq 1 +
\alpha(\tau)$ and $\alpha(0) = \alpha(T)=0$, we see that:
\begin{eqnarray}
e^{-S[x(\tau)]} &=& \frac{1}{\mathcal N} \, \int  {\mathcal D}g \, {\rm
tr}\big[g^{-1}(T) \, g(0)\big] \nonumber\\
&\times&\big( 1 + i \int_0^T d\tau {\rm tr}[\alpha(\tau) D_\tau
\dot\chi(\tau)]\big)  e^{- i \int_0^T d\tau {\rm tr}[g^{-1}(\tau)
\partial_\tau
 g(\tau) {\dot \chi}(\tau)]} \;,
\end{eqnarray}
to first order in $\alpha$. Since the term independent of $\alpha$ equals
the left hand side of the equation, we
obtain:
\begin{eqnarray}
 0 & = & \int\,  
{\mathcal D}g \, {\rm tr}\big[g^{-1}(T) \, g(0)\big]
\nonumber \\
& \times &  D_\tau \dot \chi(\tau) \; e^{- i \int_0^T d\tau {\rm
tr}[g^{-1}(\tau) \partial_\tau
 g(\tau) {\dot \chi}(\tau)]}\;.
\end{eqnarray}
Of course, this means that the saddle point equation holds on the
average. It may be regarded also as a Ward identity for the path
integral, which can be written in components as follows:
\begin{eqnarray}
 0 & = & \int\,  
{\mathcal D}g \, {\rm tr}\big[g^{-1}(T) \, g(0)\big]
\nonumber \\
& \times &  
\Big[ \ddot x_\mu(\tau) \,+\, \epsilon_{\mu \nu \rho} \,l_\nu(\tau) \, \dot
x_\rho(\tau) \Big]\; 
e^{- i \int_0^T d\tau {\rm tr}[g^{-1}(\tau) \partial_\tau
 g(\tau) {\dot \chi}(\tau)]}\;.
\end{eqnarray}
 Note that since $x_\mu(\tau)$ and $\ddot x_\mu(\tau)$ are `external'
they can be extracted out of the integral, so that we have .
\begin{equation}
\ddot x_\mu(\tau) \,+\, \epsilon_{\mu \nu \rho} \,
\langle l_\nu(\tau)\rangle \, \dot x_\rho(\tau)\;=\; 0 \;,
\end{equation}
where the average symbol means:
\begin{equation}
\langle \ldots \rangle \;\equiv \; 
\frac{\int\,{\mathcal D}g \, {\rm tr}\big[g^{-1}(T) \, g(0)\big]
\ldots e^{- i \int_0^T d\tau {\rm tr}[g^{-1}(\tau) \partial_\tau
 g(\tau) {\dot \chi}(\tau)]}}{\int\,{\mathcal D}g \, {\rm tr}\big[g^{-1}(T)
\, g(0)\big] e^{- i \int_0^T d\tau {\rm tr}[g^{-1}(\tau) \partial_\tau
 g(\tau) {\dot \chi}(\tau)]}}\;.
\end{equation}

Contracting the previous equations with $x_\mu(\tau)$, we derive the
consistency condition for the Ward identity to be valid:
\begin{equation}
\ddot x(\tau) \cdot \dot x(\tau) = 0 \;,
\end{equation}
i.e., $|\dot x(\tau)|$ must be constant, as when considering the classical
limit. 
The previous equation means that the path integral over $g$ is
not even defined when $|\dot x(\tau)|$ is not constant, since the
Ward identity (related to the invariance of the measure) would then be
violated.
Hence, the kinematical constraint on $x_\mu(\tau)$ follows in our
representation from the underlying symmetry of the gauge-group path
integral. We can then conclude that there should be a functional
$\delta$ factor imposing the previous constraint in $S$. 
Namely
\begin{equation}
e^{-S[x(\tau)]} \;=\; \delta[ \dot x^2(\tau) - v^2] \,
e^{-{\tilde S}[x(\tau)]} \;, 
\end{equation}
where $v$ is a constant and ${\tilde S}$ denotes $S$ evaluated for 
$|\dot x|=v$. We shall show at the end of this subsection that in fact $v =
1$, as in the saddle point contribution.

Let as first clean up things a bit by performing a finite change of
variables that will decouple a spin contribution
\begin{equation}\label{eq:ch1}
g(\tau) \;\equiv \; g(\tau) \, h(\tau) \;,
\end{equation}
where $h(\tau) \in SU(2)$ is a fixed ($g$-independent) element whose form
shall be determined below. The group measure is invariant:
\begin{equation}\label{eq:ch2}
{\mathcal D} ( g \, h)  \;=\; {\mathcal D} g\;,
\end{equation}
but all the other ingredients in the path integral do change. Indeed,
$$
{\rm tr}\big[g^{-1}(T) g(0)\big] \;\to\; {\rm tr}\big[g^{-1}(T) g(0)
h(0) h^{-1}(T) \big] \;,
$$
\begin{equation}\label{eq:ch21}
g^{-1} \partial_\tau g \;\to\; h^{-1} (g^{-1} \partial_\tau g ) h
\,+\, h^{-1} \partial_\tau h \;.  
\end{equation}
Thus we may write:
\begin{eqnarray}
 e^{- {\tilde S}[x(\tau)]} & = &  \frac{1}{\mathcal N} \; 
e^{- i \int_0^T d\tau {\rm tr}[h^{-1}(\tau) \partial_\tau h(\tau)
    {\dot \chi}(\tau)]} \nonumber \\
&\times & \int\,{\mathcal D}g \; {\rm tr}\big[g^{-1}(T) g(0) h(0)
  h^{-1}(T) \big] \nonumber \\
& \times &  e^{- i \int_0^T d\tau {\rm tr}[g^{-1}(\tau)\partial_\tau
    g(\tau) {\dot \chi}^{h^{-1}} (\tau)]}\;, 
\label{eq:ch3}
\end{eqnarray}
where:
\begin{equation}\label{eq:ch4}
{\dot \chi}^{h^{-1}} (\tau) \;=\; h(\tau) {\dot \chi}(\tau)
h^{-1}(\tau) \;, 
\end{equation}
and we have extracted out of the integral a factor that is independent
of $g$. The modulus of $\dot x$ is (implicitly) assumed to be constant.

Although (\ref{eq:ch3}) is valid for any function $h$,  we shall use one
that decouples spin from kinetic energy. Since the latter should only
depend on the modulus of $\dot{x}_\mu$,  the obvious choice is to use an
$h(\tau)$ such that ${\dot \chi}^{h^{-1}} (\tau)$ (\ref{eq:ch4}) is
diagonal. Since $\dot{\chi}$ is anti-Hermitian and traceless, it can be
diagonalized by an $SU(2)$ similarity transformation to a matrix
proportional to $\lambda_3$
\begin{equation}\label{eq:ch5}
{\dot \chi}^{h_{\dot \chi}^{-1}} (\tau) \;=\; \lambda_3 \, v 
\end{equation}
(where the eigenvalues of ${\dot \chi}(\tau)$ are of course $\pm
\frac{1}{2 i} v$).  $h_{\dot{\chi}}$ above is a
function of the direction of ${\dot x}(\tau)$, i.e., of the tangent
vector $t_\mu$, whose form we shall consider soon.

We know that the $x_\mu(\tau)$ paths we need to consider are
periodic in $T$. Let us also assume that $\dot{x}_\mu(T)$ points in the same
direction as  $\dot{x}_\mu(0)$, so that the tangent vectors are parallel
and pointing in the same direction.
Then \mbox{$h_{\dot{\chi}}(0)\,h_{\dot{\chi}}^{-1}(T) = 1$}, and we obtain:
\begin{eqnarray}
 e^{- {\tilde S}[x(\tau)]} & = &  \frac{1}{\mathcal N} \; 
e^{- i \int_0^T d\tau {\rm tr}[h_{\dot{\chi}}^{-1}(\tau) \partial_\tau
    h_{\dot{\chi}}(\tau) {\dot \chi}(\tau)]} \nonumber \\
& \times &  \int\,{\mathcal D}g \; {\rm tr}\big[g^{-1}(T) g(0)\big] 
\nonumber \\
& \times &  e^{- i \int_0^T d\tau {\rm tr}[g^{-1}(\tau)\partial_\tau
    g(\tau) \lambda_3 v]}\;, 
\label{eq:ch6}
\end{eqnarray}
or, introducing components in the algebra,
\begin{eqnarray}
 e^{- {\tilde S}[x(\tau)]} & = &  \frac{1}{\mathcal N} \; 
e^{\frac{i}{2} \int_0^T d\tau R_\mu(\tau) {\dot x}_\mu(\tau) }
\nonumber \\
& \times &  \int\,{\mathcal D}g \; {\rm tr}\big[g^{-1}(T) g(0)\big] 
\nonumber \\
& \times &  e^{\frac{i}{2} \int_0^T d\tau  (g^{-1}(\tau)\partial_\tau
  g(\tau))_3 v}\;,
\label{eq:ch7}
\end{eqnarray}
where:
\begin{equation}
R_\mu (\tau) \;\equiv\; \big( h_{\dot{\chi}}^{-1}(\tau) \partial_\tau
h_{\dot{\chi}}(\tau) \big)_\mu \;, 
\end{equation}
is an $\dot{x}_\mu(\tau)$ dependent vector field.

Note that we have managed to decompose the action ${\tilde S}$ into two
contributions:
\begin{equation}
{\tilde S}[x(\tau)]\;=\; S_k(v) \,+\, S_h[x(\tau)] 
\end{equation}
where $S_k$ and $S_h$ denote kinetic and spin parts, respectively. Note
that the kinetic term is just a function (not a functional).
They are defined by:
\begin{eqnarray}
e^{-S_k(v)} & = &  \frac{1}{\mathcal N} \;
\int\,{\mathcal D}g \; {\rm tr}\big[g^{-1}(T) g(0)\big] \nonumber \\
& \times &  e^{\frac{i}{2} \int_0^T d\tau (g^{-1}(\tau)\partial_\tau
  g(\tau))_3  v }\;,
\label{eq:sk}
\end{eqnarray}
and
\begin{equation}\label{eq:sh}
S_h[x(\tau)] \;=\; - \frac{i}{2} \int_0^T d\tau R_\mu(\tau) {\dot
  x}_\mu(\tau) \;. 
\end{equation}

Let us first find a more explicit form for the spin term. To that end, we note that,
${\dot x}_\mu^{h^{-1}}$, the transformed of ${\dot x}_\mu$, differs from it 
by a local (i.e., time-dependent) rotation. 
Introducing spherical coordinates for the vector $\dot{x}_\mu$: 
\begin{equation}
\dot{x}_\mu \,=\, v \, \big(\sin\theta \,\cos\phi ,\,\sin\theta
\,\sin\phi,\, \cos\theta  \big)
\end{equation}
we easily find the form 
of the matrix $h_{\dot{\chi}}(\tau)$ to be:
\begin{equation}
h_{\dot{\chi}} \,=\, \left( 
\begin{array}{cc}
\cos\frac{\theta}{2} e^{i \frac{\phi}{2}} & \sin\frac{\theta}{2}
e^{-i \frac{\phi}{2}}\\ 
- \sin\frac{\theta}{2} e^{i \frac{\phi}{2}} & \cos\frac{\theta}{2}
e^{-i \frac{\phi}{2}} 
\end{array}
  \right) \;.
\end{equation}
Finally, we obtain the three components of $R_\mu$:
\begin{equation}
R_1 \,=\,  \dot{\theta} \, \sin \phi \;, \;\; 
R_2 \,=\, - \dot{\theta} \, \cos \phi\;, \;\; 
R_3 \,=\, - \dot{\phi} \;. 
\end{equation}
Inserting this into the expression for the spin term, we find the rather
simple result:
\begin{equation}\label{eq:sh1}
S_h[x(\tau)] \;=\; \frac{i}{2} v \int_0^T d\tau \;
\cos\theta \, \dot{\phi} \;,
\end{equation}
or,  
\begin{equation}\label{eq:sh2}
S_h[x(\tau)] \;=\; \frac{i}{2} v \int d\phi \;\cos\theta \,.
\end{equation}
It is worth noting that, when $v=1$ (which will turn out to be
the case), (\ref{eq:sh2}) is a topological term, 
whose form agrees exactly with the known proposals
for the quantization of a spin-$\frac{1}{2}$ degree of freedom \cite{spin}. The
integral of the $1$-form $d\varphi \, \cos\theta$ might be converted
into the surface integral of a $2$-form which is, however,
multivalued. 

We now show that $v=1$ and calculate the kinetic term, $S_k[v]$.
We already know that $|\dot x(\tau)|$ is constant, and  by
applying a procedure which is also based on performing a change of
variables we can show that it has to be quantized, assuming any (positive)
integer value. 

We now consider in (\ref{eq:sk}) the change of variables:
\begin{equation}
g(\tau) \;\to \; g(\tau) \; h_z(\tau) \;,
\end{equation} 
where $h_z(\tau)$ is an element of $SU(2)$ with the special form:
\begin{equation}
h_z(\tau) \;=\; e^{ \frac{i}{2} \, \theta(\tau) \, \sigma_3} \;, 
\end{equation}
such that $\theta(T) - \theta(0) = 4 \pi l$, with $l \in {\mathbb Z}$.
Due to this boundary condition, we see that:
\begin{equation}
{\rm tr}\big[g^{-1}(T) g(0)\big] \; \to \;{\rm tr}\big[g^{-1}(T) g(0)\big] 
\end{equation}
and
$$
e^{\frac{i}{2} v \int_0^T d\tau (g^{-1}(\tau)\partial_\tau g(\tau))_3}
\;\to\; 
$$
\begin{equation}
e^{\frac{i}{2} v \int_0^T d\tau (g^{-1}(\tau)\partial_\tau g(\tau))_3} 
\;\times \; e^{-\frac{i}{2} v \int_0^T d\tau \partial_\tau \theta(\tau)} 
\;.
\end{equation}
Since we are just performing a change of variables in one and the same
integral, we derive the identity: 
\begin{equation}
1 \;=\; e^{-\frac{i}{2} v 
\int_0^T d\tau \partial_\tau \theta(\tau)} \;.
\end{equation}
This implies:
\begin{equation}
v \times  l \;\in\;  {\mathbb Z} \;\forall l\;.
\end{equation}
Of course, this means that $v$ must be an integer (and positive because it
is a modulus). Since, on the other hand, we know that its classical value
is $v=1$, the only possible value is $v=1$ also for the quantum case.
Then 
\begin{equation}
e^{-S_k(v)}\;=\;e^{-S_k(1)} \;=\; {\rm constant}\;.
\end{equation}

Putting together the results for kinetic and spin parts, we see that:
\begin{equation}
e^{- S[x(\tau)]} \;=\; \frac{1}{{\mathcal N}'} \, 
\delta[\dot{x}^2(\tau) - 1]\;
e^{-\frac{i}{2} \int d\phi \; \cos\theta }\,
\end{equation}
where
\begin{equation}
\frac{1}{{\mathcal N}'}\;\equiv\; \frac{e^{-S_k(1)}}{{\mathcal N}}\;,
\end{equation}
which is a function of $T$, and not of the path. Its precise form may be
written in terms of the original variables. Indeed, collecting all the
factors, we see that:
\begin{eqnarray}
\frac{1}{{\mathcal N}'}&=& 
\int {\mathcal D}p \, e^{ i \int_0^T d\tau
 p_3 (\tau)}   {\rm tr}\big[ {\mathcal P}
e^{-i \int_0^T d\tau \not
    p(\tau)} \big] \nonumber\\
&=& e^{- {\tilde S}[x(\tau)]}|_{x_\mu (\tau) \;=\; \delta_{\mu 3}}  
\;F(T)\;,
\end{eqnarray}
namely, it is determined by the action corresponding for path that
corresponds to a straight line with speed $1$ pointing in the third
direction.

We then arrive at our final expression
\begin{eqnarray}
\Gamma_f (A) & = &  \int_{0}^\infty \frac{dT}{T} \, e^{- m T} \,
F(T)\, \int_{x(0)=x(T)} {\mathcal D}x \; \delta[\dot{x}^2(\tau) - 1]
\nonumber \\ & \times & 
e^{-\frac{i}{2} \int d\phi \; \cos\theta }\,
e^{ - i e \int_0^T d\tau
{\dot x}_\mu(\tau) A_\mu[x(\tau)] }\;.
\end{eqnarray}

Regarding the fermion propagator, the same constraint on the modulus of
$\dot x$ is obtained, and the resulting expression for $D[x(\tau)]$
is:
\begin{equation}
D[x(\tau)] \;=\; \frac{1}{\mathcal N}
\delta[\dot{x}^2(\tau) - 1]\;
h_{\dot{\chi}}^{-1}(T) \; h_{\dot{\chi}}(0) 
\; e^{ - \frac{i}{2} \int_0^T d\tau \dot{\phi} \cos \theta} \;,
\end{equation}
where the matrix factors $h_{\dot{\chi}}^{-1}(T) \; h_{\dot{\chi}}(0)$
cannot be taken out of the integral over $x(\tau)$, since they depend not
only on the (fixed) boundary values for $x(\tau)$, but also on the
derivatives at the boundaries (which are not fixed).

In conclusion, we have succeeded in constructing a spacetime worldline path
integral with the spin degrees of freedom, which have been exactly
integrated in 2+1 dimensions. 
The resulting geometric picture had been qualitatively anticipated, in the
context of a superspace formulation, see e.g.~\cite{Polyakov}.

The propagator in a constant external field and the issue of the parity
violation have been computed in this first order formalism with the Migdal's
factorization \cite{us}. Our main result there was that the trace anomaly and
the related Chern--Simons current did not require any extra regularization,
and are free of any ambiguity, in contrast to other approaches. The
lengthy computation should therefore carry through in the special
parameterization without any problem, as we illustrate below with the free
propagator.  

\subsection{The free propagator}
As a simple application to test the representation, 
let us perform a derivation, of the free ($e = 0$) fermion propagator
(the free action is of course an uninteresting
constant).
$$
 G(x,y)|_{e=0} \;=\; \det(\frac{\delta_{\mu\nu}}{2}) \,\det(\partial_\tau) \; 
\int\, {\mathcal D}x \, {\mathcal D}g \, \big[g^{-1}(T) \, g(0)\big] \,
$$
\begin{equation}\label{eq:freeprop}
\times \; e^{- i \int_0^T d\tau {\rm tr}[g^{-1}\partial_\tau g(\tau) {\dot x}(\tau)]}\;.
\end{equation}

Of course, we can integrate out $x_\mu(\tau)$, what yields a $\delta$-functional
$$
\int\, {\mathcal D}x \, e^{- i \int_0^T d\tau {\rm tr}(g^{-1}\partial_\tau g{\dot x})} =\; \prod_{\mu=1}^3 \Big\{\delta [\frac{1}{2} \dot{x}_\mu(\tau)]\Big\}
$$
\begin{equation}\label{eq:fundel}
=\; \det (2 \delta_{\mu\nu})\; \prod_{\mu=1}^3 \Big\{\delta [\partial_\tau (g^{-1}\partial_\tau g)_\mu]\Big\} \;,
\end{equation}
where the factor $2$ comes from the fact that:
\begin{equation}
{\rm tr}(g^{-1}\partial_\tau g{\dot x}) = - \frac{1}{2} \; (g^{-1}\partial_\tau g)_\mu \; {\dot x}_\mu \;.
\end{equation}
Then 
$$
 G(x,y)|_{e=0} \;=\; \det(\partial_\tau) \; \int_0^\infty dT \, e^{- m T}
\, 
{\mathcal D}g \,\delta \big[\partial_\tau ( g^{-1}\partial_\tau g) \big] 
$$
\begin{equation}\label{eq:freeprop1}
\big[g^{-1}(T) \, g(0)\big] \,
e^{- i r_\mu  {(y - x)}_\mu }\;,
\end{equation}
where we have defined: 
\begin{equation}\label{eq:defr}
r \; \equiv \; g^{-1}\partial_\tau g
\end{equation}
an (antihermitian) element of $su(2)$ which can take an arbitrary constant
value, by virtue of the $\delta$-functional that appears due to the integration 
over $x_\mu(\tau)$. On the other hand, since (\ref{eq:defr}) may be solved for $g(\tau)$:
\begin{equation}
g(\tau) \;=\; g(0) \, e^{ \tau \, r} \;, 
\end{equation}
we can write:
$$
\prod_{\mu=1}^3 \Big\{\delta [\partial_\tau (g^{-1}\partial_\tau g)_\mu]\Big\}
$$
\begin{equation}
\;=\; \det^{-1} (\partial_\tau ) \; 
\int \frac{d^3 r}{(2 \pi)^3} \;\delta [ g(\tau) -  g(0) \, e^{ \tau \, r}]
\;.
\end{equation}
Using this expression we finally obtain:
\begin{equation}\label{eq:freeprop2}
 G(x,y)|_{e=0} \;=\; \int \frac{d^3 r}{(2 \pi)^3} \frac{ e^{i r \cdot (x -
y)}}{\not\! r + m}\;.  
\end{equation}

\section{Conclusions}
We have thus completed a rigorous path integral representation in the world
line of effective actions and propagators of spinors in an external gauge
field, in the conventional first time formulation and solely in terms of
ordinary (commuting) variables. The constructive proof uses gauge theory
technicalities which are also understood in terms of geometric and physical
principles. This offers a simple description of the classical limit
of the quantum spin physics from the Dirac equation, as Feynman started
looking for \cite{Fey1}. 
Beyond the conceptual interest, our formulation have practical applications
for discretizations in general and specially in the world line. In this
formalism progress has been achieved in non-perturbative computations, like
quenched all order results \cite{Nos}, pair creation and Casimir energies,
but restricted to scalar cases or very simple(one dimensional) external
fields~\cite{P1,P2}. Of course, the method has to be extended to four
dimensions~\cite{wip} which is more complicated as the functional integral
corresponds to
a hypersurface and involves groups with more Cartan elements. Our methods
should be complementary to the second order attempts, where instead of the
modulus 1 velocity (which can be related to reparametrization invariance),
the world lines velocities are Gaussian distributed with special properties
\cite{H2}.

\newpage
\section*{Appendix A: Grassmann variables}
Grassmann variables can be introduced to get an alternative expression to
the integral over gauge group variables. Indeed, one may recall, for
example, the approach introduced by Kleinert~\cite{kleinert}, to represent
the evolution operator for a spin-$\frac{1}{2}$ variable in a time
dependent magnetic field background ${\mathbf B}(\tau)$. By introducing
three real Grassmann variables $\theta_\mu$, and
performing the identifications: $B_\mu (\tau)\equiv - 2 \, p_\mu (\tau)$, 
($\mu = 1, 2, 3$) we see that:
\begin{equation}
e^{-S[x(\tau)]} \;=\; \int_{\theta(T) = - \theta(0)} \,  
{\mathcal D}p \; {\mathcal D}\theta \, 
e^{-\int_0^T d\tau \big[\frac{1}{4} \theta_\mu {\dot \theta}_\mu \,-\, 
\frac{1}{2}\epsilon_{\nu\nu\lambda} p_\mu \theta_\nu \theta_\lambda
\,-\, i \, p_\mu {\dot x}_\mu \big]}\;,
\end{equation}  
where the antiperiodicity for the Grassmann variables is necessary in order
to get the trace of the path-ordered exponential. The other object,
$D[x(\tau)]$ is instead given by 
\begin{equation}
D[x(\tau)] \;=\; \int_{\theta_\mu(0) = \theta_\mu^{(i)}}^{\theta_\mu(T) =  \theta_\mu^{(f)}}  \,  {\mathcal D}p \; {\mathcal D}\theta \, 
e^{-\int_0^T d\tau \big[\frac{1}{4} \theta_\mu {\dot \theta}_\mu \,-
\,\frac{1}{2}
\epsilon_{\nu\nu\lambda} p_\mu \theta_\nu \theta_\lambda 
\,-\, i \, p_\mu {\dot x}_\mu \big]}\;,
\end{equation} 
where the boundary values for $\theta$ ($\theta^{(i)}$ and $\theta^{(f)}$)
can be chosen in order to evaluate the desired matrix element of the
fermion propagator, by selecting the appropriate coherent states at the
initial and final times.

By proceeding with the analogy with the system corresponding to a spin
in a magnetic field~\cite{kleinert}, an interesting conclusion may be drawn
from the (quantum) equations of motion for the corresponding operators. 
The relevant dynamical operators are the `spin' ${\hat S}_\mu$
\begin{equation}
{\hat S}_\mu \;\equiv \;-\frac{i}{4} \varepsilon_{\mu\nu\lambda} \theta_\nu
\theta_\lambda, 
\end{equation}
and the velocity operator $\hat{\dot x}_\mu$. The equations
of motion for the spin degrees of freedom are:
\begin{equation}
{\dot{\hat S}}_\mu(\tau) \;=\; - \epsilon_{\mu\nu\lambda} \,\pi_\nu (\tau) 
\, {\hat S}_\lambda(\tau), 
\end{equation}
and the constraint equation for $p_\mu$, which is a Lagrange multiplier:
\begin{equation}
\dot{\hat{x}}_\mu(\tau) \;=\; 2 {\hat S}_\mu(\tau)\;. 
\end{equation}
Then we see that:
\begin{equation}
{\dot{\hat x}}_\mu(\tau) \;=\; - \epsilon_{\mu\nu\lambda} \,\pi_\nu (\tau) 
\, {\hat x}_\lambda(\tau) \;,
\end{equation}
consistent with the semiclassical equations.
\newpage
\section*{Acknowledgements}
We thank Orlando Alvarez and Juan Luis Ma\~nes for suggestions and for
careful reading of  the manuscript. 
J.S.-G. and R.A.V. thank MCyT (Spain) and FEDER (FPA2005-01963), and
Incentivos from Xunta de Galicia. 
C.D.F. has been partially supported by CONICET. He acknowledges the kind
hospitality of the members of Facultade de F\'{\i}sica, USC, where part of
this work has been done.

\end{document}